\documentclass[preprint]{aastex}

\def\etal{{\rm et al.\ }}

\def\log {{\rm log}}

\def\lsim{\mathrel{\hbox{\rlap{\hbox{\lower4pt\hbox{$\sim$}}}\hbox{$<$}}}}
\def\gsim{\mathrel{\hbox{\rlap{\hbox{\lower4pt\hbox{$\sim$}}}\hbox{$>$}}}}

\begin{document}

\title{The Phase-Space Density Profiles of Cold Dark Matter Halos}

\author{James E. Taylor and  Julio F. Navarro\altaffilmark{1}}

\affil{Department of Physics and Astronomy \\ University of Victoria, 
 Victoria, BC, V8P 1A1, Canada}

\altaffiltext{1}{CIAR Scholar and Alfred P. Sloan Research Fellow}

%\lefthead{Taylor \& Navarro}
%\righthead{Phase-Space Density of CDM Halos}\author{}

\begin{abstract}
We examine the coarse-grained phase-space density profiles of a set of recent,
high-resolution simulations of galaxy-sized Cold Dark Matter (CDM) halos. Over
two and a half decades in radius the phase-space density closely follows a
power-law, $\rho/\sigma^3 \propto r^{-\alpha}$, with $\alpha \approx
1.875$. This behaviour closely matches the self-similar solution obtained by
Bertschinger for secondary infall of gas onto a point-mass perturber in a
uniformly expanding universe. On the other hand, the density profile
corresponding to Bertschinger's solution (a power-law of slope $r^{2\alpha-6}$)
differs significantly from the density profiles of CDM halos. CDM halo density
profiles are clearly not power laws, and have logarithmic slopes that gradually
steepen with radius, roughly as described by Navarro, Frenk \& White (NFW). We
show that isotropic, spherically-symmetric equilibrium mass distributions with
power-law phase-space density profiles form a one-parameter family of structures
controlled by the ratio of the local velocity dispersion to the ``natural''
velocity dispersion at some fiducial radius, $r_0$; $\kappa=4\pi G \rho(r_0)
r_0^2/\sigma(r_0)^2$. For $\kappa=\alpha=1.875$ one recovers the power-law
solution $\rho \propto r^{2\alpha-6}$. As $\kappa$ increases, the density
profiles become quite complex but still diverge like $r^{2\alpha-6}$ near the
center. For $\kappa$ larger than some critical value, $\kappa_{\rm
crit}(\alpha)$, solutions become non-physical, leading to negative densities
near the center. The critical solution, $\kappa =\kappa_{\rm crit}$, corresponds
to the case where the phase-space density distribution is the narrowest
compatible with the power-law phase-space density stratification
constraint. Over three decades in radius the critical solution is
indistinguishable from an NFW profile, although its logarithmic slope
asymptotically approaches $-2\alpha/5 = -0.75$ (rather than $-1$) at very small
radii. Our results thus suggest that the NFW profile is the result of a
hierarchical assembly process that preserves the phase-space stratification of
Bertschinger's spherical infall model but which ``mixes'' the system maximally,
perhaps as a result of repeated merging, leading to a relatively uniform phase-space
density distribution across the system. This finding offers intriguing clues as
to the origin of the similarity in the structure of dark matter halos formed in
hierarchically clustering universes.
\end{abstract}


\section{Introduction}

In the currently favored paradigm for the formation of structure in the
Universe, luminous objects such as galaxies and galaxy clusters are embedded
within extended halos of cold dark matter (CDM). The non-linear equilibrium
structure of these halos has long been thought to contain clues as to the
astrophysical nature of dark matter, a fact that has fueled strong interest in
detailed theoretical predictions for the structure of dark matter halos, as well
as in constraints placed by observations of disk galaxy rotation curves (Frenk
\etal 1988; Flores \etal 1993; Flores \& Primack 1994; Moore 1994; Moore \etal
1999b, McGaugh \& de Blok 1998, van den Bosch \etal 2000; Swaters, Madore \&
Trewhella 2000; van den Bosch \& Swaters 2000), by gravitational lensing of
galaxies and clusters (see, e.g., Tyson, Kochanski \& dell'Antonio 1998,
Williams, Navarro \& Bartelmann 1999), and by detailed studies of the stellar
dynamics of spheroidal galaxies (Carollo \etal 1995; Rix \etal 1997; Gerhard
\etal 1998; Cretton \etal 2000; Kronawitter \etal 2000).

Unfortunately, observational constraints tend to be strongest near the center
of dark matter halos -- where most of the luminous material in galaxies resides
-- but also where theoretical predictions are least robust. This is because of
the difficulties inherent to simulating accurately through N-body methods
regions where overdensities exceed $\sim 10^6$ and where particles may have
completed thousands of orbits during a Hubble time. In spite of these
difficulties, there has been steady progress in our theoretical understanding of
the non-linear structure of virialized dark matter halos, driven largely by
direct numerical simulation. Following on pioneering numerical studies by Quinn,
Salmon \& Zurek (1986), Frenk \etal (1988), Dubinski \& Carlberg (1991), and
Crone, Evrard \& Richstone (1994), Navarro, Frenk \& White (1996, 1997,
hereafter NFW) found that, regardless of mass and of the value of the
cosmological parameters, the density profiles of dark matter halos formed in
various hierarchical clustering cosmogonies were strikingly similar. This
``universal'' structure can be characterized by a spherically-averaged density
profile which differs substantially from the simple power-laws, $\rho(r) \propto
r^{-\beta}$, predicted by early theoretical studies (Gunn \& Gott 1972, Fillmore
\& Goldreich 1984, Hoffmann \& Shaham 1985, White \& Zaritsky 1992). The profile
steepens monotonically with radius, with logarithmic slopes shallower than
isothermal (i.e. $\beta < 2$) near the center, but steeper than isothermal
($\beta>2$) near the virial radius of the system.

The NFW result has been confirmed by a number of subsequent studies (Cole \&
Lacey 1996, Huss, Jain \& Steinmetz 1999, Fukushige \& Makino 1997, Moore \etal
1998, Jing \& Suto 2000), although there is some disagreement amongst authors
regarding the innermost value of the logarithmic slope. NFW argued that a
fitting formula where $\beta=(1+3y)/(1+y)$ (where $y=r/r_s$ is the radius in
units of a suitably defined ``scale radius'' $r_s$) provides a very good fit to
the density profiles of simulated halos over two decades in radius. Moore \etal
(1998), Ghigna \etal (2000), and Fukushige \& Makino (2000) have argued that
$\beta$ converges to a value of $\sim 1.5$ near the center,
rather than the $1$ expected from the NFW fit. Kravtsov \etal (1998) initially
obtained much shallower inner slopes ($\beta \sim 0.7$) in their numerical
simulations, but have now revised their conclusions; these authors now argue
that CDM halos have steeply divergent density profiles but, depending on
evolutionary details, the slope of a galaxy-sized halo at the innermost resolved
radius may very between $-1.0$ and $-1.5$ (Klypin \etal 2000). 

Steep inner slopes have been traditionally disfavored by rotation curve data, a
fact that has often been used to ``rule out'' CDM as a viable cosmogony (Flores
\& Primack 1994, Moore 1994, McGaugh \& de Blok 1998, Blais-Ouellette \etal 2001, 
C{\^o}t{\'e} \etal 2000, de Blok \etal 2001). However, recent reanalysis of the
available data suggests that most rotation curves are broadly consistent with
``cuspy'' dark matter cores, provided that the innermost slope is shallower than
$\beta \sim 1.5$ and that the ``concentration'' of the halos is low (Navarro
1998, van den Bosch \etal 2000, van den Bosch \& Swaters 2000). 

Although there may not be at present broad consensus regarding how steep the
innermost slope is, or even whether there should be a well-defined asymptotic
innermost slope, there is agreement that it will take extraordinary
computational effort to reach a robust resolution of the controversy. What is
required is a statistically significant sample of galaxy-sized halos with
sub-kpc resolution, an extremely onerous computational task that will stretch
the capabilities of the most powerful massively parallel computers. Steps in
this direction are currently being taken (e.g., Power et al., in preparation),
but it will take some time until these efforts yield conclusive results.

From the theoretical point of view, a number of plausible arguments have been
advanced in order to try and explain the innermost behaviour of dark matter
density profiles from stellar dynamical principles. These efforts, however, tend
to give non-unique results and have so far been unable to explain the remarkable
similarity in the structure of dark matter halos of widely different mass formed
in a variety of cosmogonies (Evans \& Collett 1997, Syer \& White 1998, Nusser \& Sheth 1999, {\L}okas \& Hoffman 2000).  In this paper we investigate an empirical
alternative to analytic efforts addressed at estimating the innermost slope of
the density profile. Our proposal exploits the similarity between the
phase-space density profiles of CDM halos and that of the self-similar solution for
spherical collapse in an expanding universe found by Bertschinger (1985). In
addition, this offers an attractive scenario for understanding the shape of halo
density profiles as well as a powerful tool for estimating their slope near the
center.

The outline of this paper is as follows. In Section 2, we present the
phase-space density profiles for galaxy-sized CDM halos in several recent
simulations, and show that they are well approximated, for over two decades in
radius, by a power law.  In Section 3, we investigate equilibrium density
profiles consistent with this constraint by solving the Jeans equation for
spherically symmetric systems with isotropic velocity dispersion
tensors. Section 4 compares these results with the results of numerical
simulations. We discuss and summarize our findings in Section 5.

\section{The Phase-Space Density Profile of CDM Halos}

We have chosen for our analysis three high-resolution simulations of
galaxy-sized CDM halos identified at $z \sim 0$. Each of these simulations has
of order $10^6$ particles within the virial radius of the halo, and was run
with PKDGRAV, a massively-parallel N-body code developed by Joachim Stadel and
Thomas Quinn at the University of Washington. These simulations are amongst the
largest carried out so far for galaxy-sized objects, and represent a major
investment in computational resources.

Two of the halos were run in the former ``standard'' CDM cosmology (SCDM,
$\Omega_m = 1, \Lambda = 0, h = 0.5, \sigma_8=0.7$) and have circular velocities
of $\sim 180$ and $\sim 160$ km s$^{-1}$, respectively. These circular
velocities are measured at the ``virial'' radius, $r_{200}$, where the mean
inner density of the system is $200$ times the critical value for closure. These
two halos are part of the ``Local Group'' simulation reported in Moore \etal
(1999a). The third halo has a circular velocity of $200$ km s$^{-1}$ and was run
in the currently popular LCDM cosmogony ($\Omega_m = 0.3, \Lambda = 0.7, h =
0.65, \sigma_8=0.9$). Gravitational softenings were chosen to be $\sim 3\times
10^{-3}$ and $1.5 \times 10^{-3} \ r_{200}$, for the SCDM and LCDM runs,
respectively. PKDGRAV uses a multi-stepping algorithm to integrate the equations
of motion; particles with the smallest time bins may take up to $100,000$
timesteps to evolve from the initial redshift ($z_i=99$ for SCDM and $z_i=49$
for LCDM) to the present. The initial conditions for the SCDM runs were setup
using the algorithms devised by the N-body shop at the University of
Washington. Those for the LCDM run were set up in a completely independent way
using the algorithms described by Navarro, Frenk \& White (1996, 1997; see also
Efstathiou \etal 1985). As we describe below, the results of these two
independent runs are consistent with each other, which suggests that our
conclusions are independent of both the cosmological model adopted and of the
particular choice of initial conditions setup algorithms.

Figure 1 shows the spherically-averaged phase-space density profile of the three
CDM halos. Solid (dashed) lines are used for the SCDM (LCDM) halos. The
phase-space density is computed in spherical bins containing $2,000$ particles
each and is defined as $\rho/\sigma^3$, where $\rho$ is the mass density and
$\sigma$ is the 1-D velocity dispersion in the bin.\footnote{The velocity
dispersion tensor is roughly isotropic near the center and only mildly radially
biased in the outer regions.} For ease of comparison, we have chosen to
normalize $\rho/\sigma^3$ so that all three curves coincide at $0.01 \
r_{200}$. The important point illustrated by this figure is that, over more than
two decades in radius, the phase-space density profile is very well approximated
by a power law of slope $-1.875$ (thin straight line). This is quite
remarkable, given that both the density profiles (shown in Figure 2) and the
velocity dispersion profiles of these halos deviate quite strongly from simple
power laws, as described by NFW.

Also remarkable is that the slope of this power-law coincides with the
self-similar solution derived by Bertschinger (1985) for secondary infall onto a
spherical perturbation in an unperturbed Einstein-de Sitter universe:
Bertschinger's solution is plotted with solid circles in Figure 1. This solution
corresponds to the self-similar equilibrium configuration of a $\gamma=5/3$ gas
formed by spherical accretion onto a point-mass perturber in an otherwise
uniform Einstein-de Sitter universe. The quantity shown by the solid circles is
the quantity ``equivalent'' to the phase-space density, $\rho^{5/2}/P^{3/2}$,
where $P$ is the local (isotropic) pressure. As discussed by Bertschinger, this
solution is the most appropriate to compare with our numerical results for CDM
halos, given that the velocity dispersion tensor in this case is only mildly
anisotropic. Radii are normalized assuming that $r_{200}$ equals the shock
radius of the self-similar solution, which corresponds to roughly one-third of
the turnaround radius. The vertical normalization is arbitrary and has been
chosen to match the N-body results at 0.01\,$r_{200}$. 
Taking $\rho^{5/2}/P^{3/2} \propto
(T/\rho^{\gamma-1})^{-5/2\gamma}$ to be a measure of the local ``entropy'' of
the system, Figure 1 shows that CDM halos have the same radial entropy
stratification as the simple spherical collapse solution. It is possible that
this power-law stratification is a fundamental property which underlies the
similarity of structure of cold dark matter halos.

\section{Density Profiles}

Density profiles consistent with the power-law phase-space density profile shown
in Figure 1 can be obtained by assuming hydrostatic equilibrium. For an
isotropic, spherically-symmetric system of collisionless particles, the Jeans
equation may be written as
\begin{equation}\label{Jeans}
{{d(\rho\,\sigma^2)}\over{dr}} = -\rho\ {{d\Phi}\over{dr}} =
-\rho{{GM(<r)}\over{r^2}}\ ,
\end{equation}
where $\Phi$ is the gravitational potential, and $M(<r)$ is the mass interior to
$r$ (Binney and Tremaine, 1987, p.\ 198). Equation (\ref{Jeans}) is equivalent to
the equation of hydrostatic equilibrium for a gas of pressure $P =
\rho\,\sigma^2$. Dividing both sides of the equation by $-G\rho/r$, and taking
derivatives with respect to $r$, we can rewrite this equation as,
\begin{equation}\label{J2}
{{d}\over{dr}}\left({{-r^2}\over{G\rho}}\left({{d(\rho\,\sigma^2)}\over{dr}}\right)\right)
= {{d}\over{dr}}M(< r) = 4\pi\rho\,r^2,
\end{equation}
where the last equivalence applies to a self-gravitating system.  Assuming that
the phase-space density is a power-law of radius, i.e. that
\begin{equation}\label{PL1}
\rho/\sigma^3 (r) = (\rho_0/\sigma^3_0)(r/r_0)^{-\alpha},
\end{equation}
where $r_0$ is an (arbitrary) reference radius and $\sigma_0 = \sigma(r_0)$ and
$\rho_0 = \rho(r_0)$, and defining the dimensionless variables, $x \equiv r/r_0$
and $y \equiv \rho/\rho_0$, (\ref{J2}) can then be written as,
\begin{equation}\label{J5}
{{d}\over{dx}} \left( {{-x^2}\over{y}} {{d}\over{dx}}
\left(y^{5/3}x^{2\alpha/3}\right)\right) = \kappa y\,x^2,
\end{equation}
where
\begin{equation}
\kappa \equiv {{4\pi G \rho_0 r_0^2}\over{\sigma_0^2}} 
\end{equation}
is a dimensionless measure of the velocity dispersion at $r_0$. A simple
interpretation of the parameter $\kappa$ can be found by choosing $r_0$ to be
$r_p$, the radius where the circular velocity of the system peaks. In this case,
and since $\rho(r_p) = M(<r_p)/4\pi r_p^3$, we can rewrite $\kappa$ as,
\begin{equation}\label{J5a}
\kappa = {{4\pi G \rho_p r_p^2}\over{\sigma_p^2}} = {{G M(<r_p)/r_p}\over{\sigma_p^2}}
= {{V_p^2}\over{\sigma_p^2}}.
\end{equation}
In other words, $\kappa$ measures the local velocity dispersion of the system in
units of its circular velocity at the radius where the circular velocity peaks.

Equation (\ref{J5}) has the form,
\begin{equation}\label{J6}
y'' + f(y',y,x) = 0,
\end{equation}
(where $y' \equiv dy/dx, y'' \equiv d^2y/dx^2$) so given initial conditions
$y(1) = \rho(r_0)/\rho_0 = 1$ and $y'(1) = y'_1$ at $x= 1 (r= r_0)$, we can
solve it numerically for $r > r_0$ or $r < r_0$ by integrating inwards or
outwards. The resulting solutions are parameterized by $\alpha$, $y'_1$ and
$\kappa$, so for $\alpha = 1.875$, we obtain a two-parameter family of
solutions. Clearly, integrating (\ref{J6}) with respect to some different
variable $z = t\,x$, where $t$ is a constant, will produce an identical
solution, provided that we scale $y'_1$ by $t$ and $\kappa$ by $t^{2\alpha/3 -
2}$, so in fact the solutions are degenerate and there is effectively a single
free parameter (which we can take to be $\kappa$) that characterizes the full
solution set.\footnote{This assumption breaks down if we choose $y'_1$ to be
positive, or negative but too small. In this case $\rho(r)$ {\it increases} with
$r$ and the solutions are unstable and lack direct physical interpretation.}

Equation (\ref{J6}) admits power-law solutions. Solving for $y = x^{-\beta}$ gives
the following conditions. For $\kappa=0$ we have,
\begin{eqnarray}
\beta &=& 2\alpha/5 \nonumber\\
{\rm and} \ \  \beta &=& \alpha + 3/2,
\end{eqnarray}
while for $\kappa \ne 0$ the conditions are,
\begin{eqnarray}
\beta &=& 6 - 2\alpha \nonumber\\
\kappa &=& 8\, (\alpha - 3/2)(5/2-\alpha).
\end{eqnarray}
For example, the singular isothermal sphere is recovered for
$\kappa=\alpha=\beta=2$.

Thus assuming $\kappa \ne 0$, that is, a finite velocity dispersion at $r_0$,
there is a single power-law solution to (\ref{J6}) with slope $\beta =6 -
2\alpha= 2.25$ for $\alpha = \kappa= 1.875$.\footnote{Curiously, besides the
singular isothermal sphere this is the only other case where the power-law
solution has $\kappa=\alpha$.} It is convenient to set $y'_1 = 2\alpha-6 =
-2.25$ and to generate the family of solutions simply by varying
$\kappa$. Figure 2 shows a few of these solutions (for $\alpha = 1.875$),
labeled by the particular value of $\kappa$ beside each curve. With the
exception of the power law, the shape of the equilibrium density profiles is
complex, with an outer cutoff, one or more inflection points, and an inner
cusp. For $\kappa$ larger than some critical value, $\kappa_{\rm crit}(\alpha)$,
the solutions become non-monotonic, and the density vanishes at a finite
radius. All solutions with $1.875<\kappa<\kappa_{\rm crit}$ appear to have a
steep inner cusp with asymptotic slope close to $\beta = 6 - 2\alpha$, but the
critical solution approaches asymptotically $\beta = 2\alpha/5 = 0.75$ as $r$
tends to $0$.

\section{Comparison with Numerical Density Profiles}

As discussed in the previous section, the isotropic Jeans equation admits a
family of solutions for the density profile under the constraint $\rho/\sigma^3
\propto r^{-\alpha}$.  The family includes a power-law, $\rho \propto
r^{-\beta}$, with $\beta=6-2\alpha=9/4=2.25$ (for $\alpha=15/8=1.875$), which
corresponds to Bertschinger's spherical infall solution. This is shown with
filled circles in Figure 3, where we also show the density profiles
corresponding to the CDM halos in the N-body simulations. Radii have been scaled
here, as in Figure 2, to the radius where the slope of the density profile is
$-2.25$, and densities to the critical density. Clearly, the power-law solution
is a poor fit to the result of the N-body simulations, which become noticeably
shallower than $r^{-9/4}$ near the center.

The N-body results can be well fitted by the profile proposed by NFW, which is
shown by the open triangles in Figure 3. Interestingly, over approximately three
decades in radius, {\it the shape of the NFW profile is essentially
indistinguishable from the ``critical'' solution} alluded to
above. Discrepancies occur only at large radii, where substructure leads to
systematic deviations from the power-law behaviour for the phase-space density.
The meaning of the critical solution may be illustrated by considering the
phase-space density distribution corresponding to the different solutions. This
is shown in Figure 4 for the solutions illustrated in Figure 2, for
systems normalized to have the same total mass, energy, and mean
density inside some fiducial radius. 
Clearly, the phase-space distribution function is broadest for the
power-law solution ($\kappa=\alpha=1.875$) and gets increasingly narrower as
$\kappa$ increases; the critical solution ($\kappa=\kappa_{\rm crit}$)
corresponds then to the most sharply peaked phase space distribution compatible
with a monotonically decreasing (non-hollow) density profile and with the entropy
stratification constraint. The critical solution may thus be
interpreted as a ``maximally mixed'' configuration where the phase-space density
is as uniform as possible across the system.

This leads to the following interpretation for the origin of the NFW
profile. Gravitational assembly of CDM halos leads to a simple power-law radial
stratification of the phase-space density. If spherical symmetry is imposed, as
in the case treated by Bertschinger (1985), the collapsing radial mass
shells generate progressively larger ``entropies'' 
(lower phase-space densities) as they pass
through the shock and settle into hydrostatic equilibrium, leading to steeply
cusped power-law profiles with slope $\beta=6-2\alpha$. On the other hand, when
the assumption of spherical symmetry is released and the collapse proceeds
through many stages of hierarchical merging, mass shells are continuously
``mixed'' and the profiles tend to the critical solution: that corresponding to
the most uniform entropy distribution compatible with a non-hollow
density profile and with the power-law entropy stratification constraint. This density
profile closely resembles the NFW profile over a large dynamic range in radius.

Finally, it is important to note that, despite the similarity shown in Figure 3,
there are important differences between the NFW profile and the critical
solution, the most notable being that the latter tends to an asymptotic central
slope of $-\beta = -2\alpha/5 = -0.75$ rather than to $-1$ as in NFW's fitting
formula.  A simple approximation to the radial dependence of the slope of the
``critical'' density profile is given by
\begin{equation}
{{d\ln \rho}\over{d \ln x}} = -{0.75 + 2.625 \, x^{1/2}\over 1+0.5 \, x^{1/2}},
\end{equation}
which is accurate to $3 \% $ for $x = r/r_0 < 4$. Here $r_0$ is the
radius where the logarithmic slope of the density profile equals $-2.25$
and $r_0 = (5/3) r_s$, in terms of the NFW scale radius $r_s$.

\section{Discussion}

The power-law phase-space density profile thus offers a natural way to
describe the structure of dark matter halos at radii where simulations
become increasingly difficult and expensive, but also where observational
constraints are strongest. Provided that the velocity dispersion tensor remains
nearly isotropic, the ``critical'' solution provides a clear prediction as to
the behaviour of the logarithmic slope of the density profile: it should become
progressively shallower towards the center, converging asymptotically to a 
value of
$-2\alpha/5=-0.75$. This is interesting since, as mentioned in \S1, slopes
shallower than $-1.5$ appear to be consistent with the recent reanalysis of the
rotation curve dataset by van den Bosch \etal (2000) and by van den Bosch \&
Swaters (2000). It is important to stress, however, that a shallow central
slope does not guarantee consistency with observations, which constrain the
detailed radial dependence of the density profile slope as well. Our results do,
however, offer a clear prediction for extrapolating the mass profiles to regions
that are very difficult to probe numerically.  Is the extrapolation of the
power-law behaviour to very small radii warranted?  This question ultimately
will have to be answered by direct numerical simulation, although there is no
obvious a priori reason why a power law scaling that is valid for over two
decades in radii should break down nearer the center.

One important point to note is that the critical solution is clearly at odds
with the proposal of Moore \etal (1998) and Ghigna \etal (2000) that the
innermost slopes of CDM halos converge to a value not shallower than about
$-1.5$. However, it should be emphasized that their conclusion was based on the
simulation of a {\it single} halo simulated in a standard CDM universe (SCDM)
and on a significantly different mass regime (galaxy clusters) than probed
here. Thus the possibility remains that this particular system may not be
representative of the general population or that the density profiles of
clusters are steeper than those of galaxy-sized halos. Since these authors use
the same N-body code as in the present work and, indeed, we use their own
results for galaxy-sized halos in this manuscript, it is unlikely that the
discrepancy is due to subtle errors associated with the numerical setup of the
simulations.  Confirming which of these possibilities holds will require a
statistically significant sample of halos simulated with resolution comparable
to the systems used here. Finally, it is also possible that the cluster
simulated by Moore \etal (1998) and Ghigna \etal (2000) differs from the
galaxy-sized halos we present here in other, more subtle ways. For example, it
may be significantly more triaxial than the systems analyzed here, or perhaps
its velocity dispersion tensor is very anisotropic, in conflict with the
assumptions of this work. Again, a detailed reanalysis of the discrepant system,
extended to a statistically meaningful sample, appears necessary in order to
explain this discrepancy conclusively. We are currently working on this issue
and plan to report our findings soon (Stadel et al., in preparation).

\section{Summary}

We examine the spherically averaged phase-space density profiles of N-body
simulations of CDM halos in the SCDM and LCDM cosmologies and find that they are
very well approximated by a power-law ($\rho/\sigma^3 \propto r^{-1.875}$) over
more than two decades in radius. The slope of this power law is consistent with
that of the self-similar solution for spherical secondary infall derived by
Bertschinger (1985). Assuming that the phase-space density profile is a power
law, and assuming isotropy, hydrostatic equilibrium is satisfied by a family of
density profiles controlled by a single-parameter. This parameter is determined
by the ratio of the velocity dispersion to the ``natural'' velocity dispersion
of the system at some fiducial radius, $r_0$, $\kappa=4 \pi G \rho(r_0)
r_0^2/\sigma(r_0)^2$. The parameter $\kappa$ can also be expressed in terms of
the ratio of the circular velocity to the 1-D velocity dispersion at $r_p$, the
radius where the circular velocity peaks ($\kappa=V(r_p)^2/\sigma(r_p)^2$). For
$\kappa=1.875$, the density profile is a power-law, which agrees with
Bertschinger's solution. As $\kappa$ increases the density profiles become
increasingly curved, although they still approach the steep power-law divergent
behaviour near the center. For $\kappa$ greater than some critical value
$\kappa_{\rm crit} \simeq 2.678$ the density profiles become unphysical, vanishing at
some finite radius near the center.

The ``critical'' density profile ($\kappa=\kappa_{\rm crit}$) corresponds to the
maximum value of $\kappa$ consistent with a non-vanishing density profile at the
center and {\it is essentially indistinguishable from the profile proposed by
Navarro, Frenk \& White} (1996, 1997) for over two decades in radius. This
solution corresponds to the most uniform (more sharply peaked) phase-space
distribution compatible with a monotonically decreasing density profile and 
with the
entropy stratification constraint. This suggests that the structure of CDM halos
is determined by a radial phase-space density stratification process similar to that
established through collapse onto a point mass perturbation in an unperturbed
expanding universe, and by the uniformization of phase-space density that occurs
presumably as a result of the many merger and satellite accretion events that
characterize the assembly of a CDM halo. This identification leaves a couple of
important questions unanswered, however: (i) why should the phase-space density be a
power law of radius?, and (ii) why is the exponent the same as in Bertschinger's
self-similar solution?. Although we have no clear answer to these questions at
this point, our results suggest that explaining the origin of the structural
similarity of CDM halos pointed out by Navarro, Frenk \& White may entail
unraveling why the radial stratification of phase-space density in CDM halos is
the same power-law generated by the simple spherical collapse model. Discovering
a mechanism that achieves this may provide a simple explanation for the
universal structure of cold dark matter halos.

\acknowledgements

The authors wish to thank Joachim Stadel and Tom Quinn for kindly making their
massively parallel code PKDGRAV available as well as for their help and guidance
running the simulations. We extend our thanks to the other members of their
N-body collaboration, B.\ Moore, F.\ Governato, and G.\ Lake, for allowing us to
use the data from their simulations. We also thank Simon White for providing
insightful comments on a preliminary version of this manuscript. The Natural
Sciences \& Engineering Research Council of Canada (NSERC) and the Canadian
Foundation for Innovation have supported this research through a postgraduate
scholarship to JET and through various grants to JFN.

% Bibliography

%Figures

\begin{figure}
\plotone{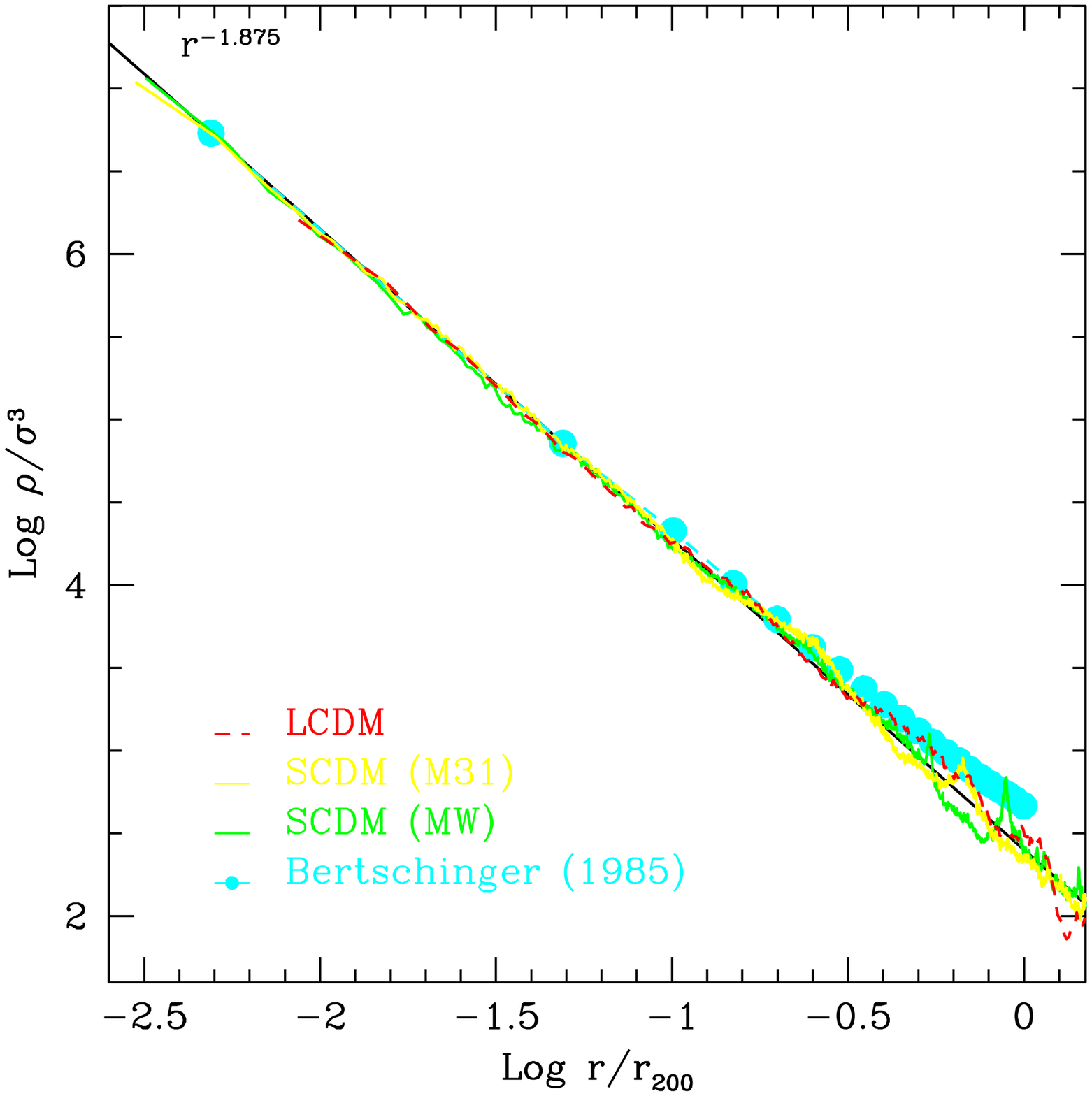}
%   \centerline{\psfig{figure=figs/wfig1.epsi,width=1.0\linewidth,clip=,angle=0}}
\caption[fig1.eps]
{The phase-space density profiles of three galaxy-sized CDM halos. Solid lines
correspond to the SCDM halos and dashed lines correspond to the LCDM
halo. Vertical normalizations are arbitrary and have been chosen so that the
curves coincide at about $0.01 \, r_{200}$. Radii are normalized to the virial
radius, $r_{200}$. The solid circles indicate the self-similar solution obtained
by Bertschinger (1985) for spherical infall of gas onto a point mass perturber
in a uniform Einstein-de Sitter universe. Radii for this solution have been
normalized by assuming that the shock radius in the solution equals $r_{200}$. A
power-law of slope $-1.875$ is shown for comparison (thin solid
line).\label{fig1}}
\end{figure}

\begin{figure}
\plotone{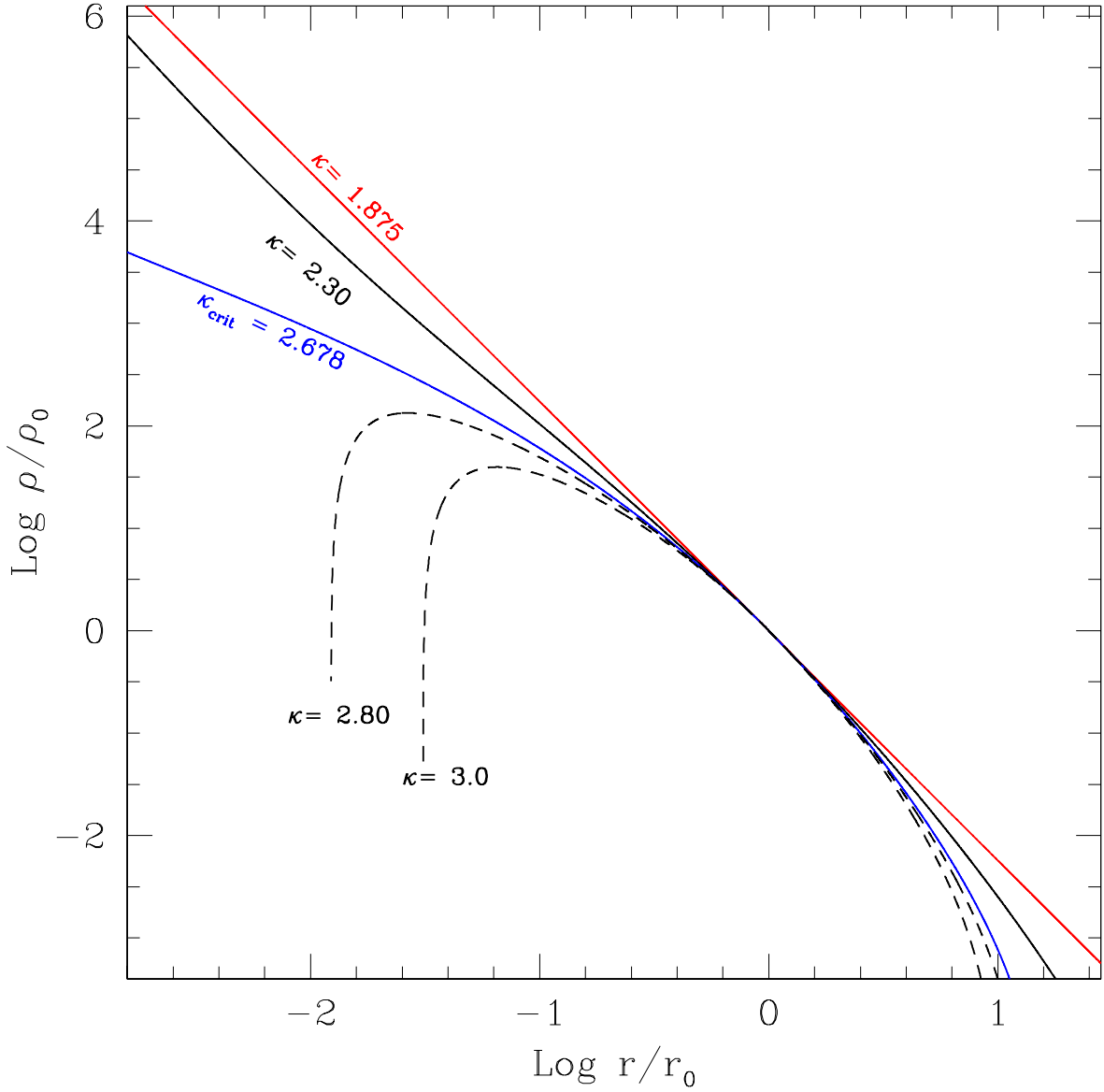}
%   \centerline{\psfig{figure=figs/wfig2a.epsi,width=1.0\linewidth,clip=,angle=0}}
\caption[fig2.eps]
{Density profile solutions to the isotropic, hydrostatic equilibrium (Jeans)
equation, for spherically symmetric systems with a power-law phase-space
density, $\rho/\sigma^3 \propto r^{-1.875}$. The curves are generated by setting
$\beta=-d(\log \rho)/d(\log r) = 2.25$ and $\rho(r_0) = \rho_0 = 1$ at some
fiducial radius, $r_0$, and by varying the free parameter $\kappa$. The value of
$\kappa$ is used to label each curve. \label{fig2}}
\end{figure}

\begin{figure}
\plotone{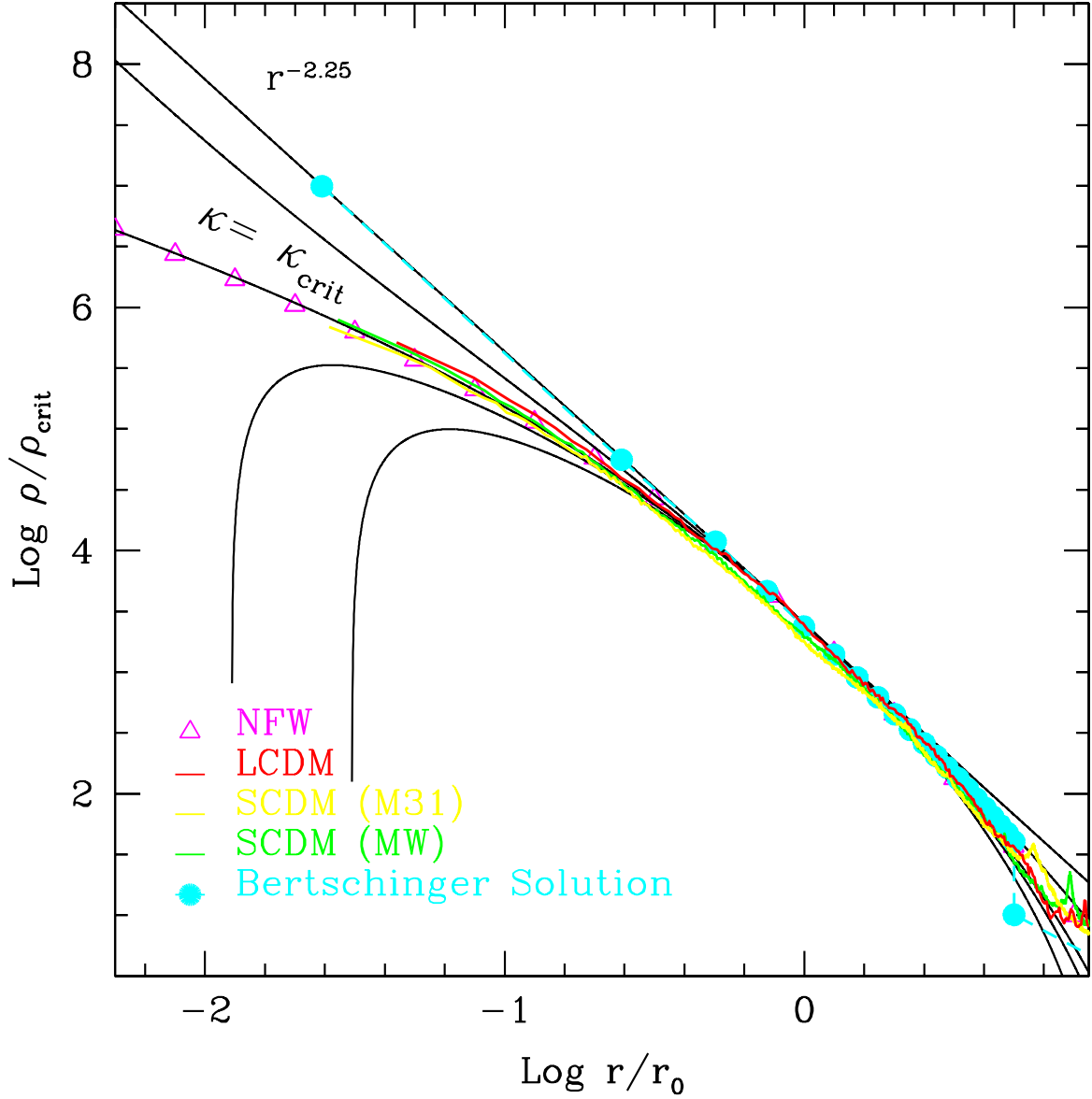}
%   \centerline{\psfig{figure=figs/wfig2.epsi,width=1.0\linewidth,clip=,angle=0}}
\caption[fig3.eps]{Density profiles of the CDM halos and the
Bertschinger solution, compared with the analytic solutions shown in Figure 2. These profiles
deviate significantly from the power-law solution ($\rho \propto r^{-2.25}$) and
from Bertschinger's solutions, but are well fitted by an NFW profile (open
triangles) and by the ``critical'' ($\kappa=\kappa_{\rm crit}$) solution (see
text).\label{fig3}}
\end{figure}

\begin{figure}
\plotone{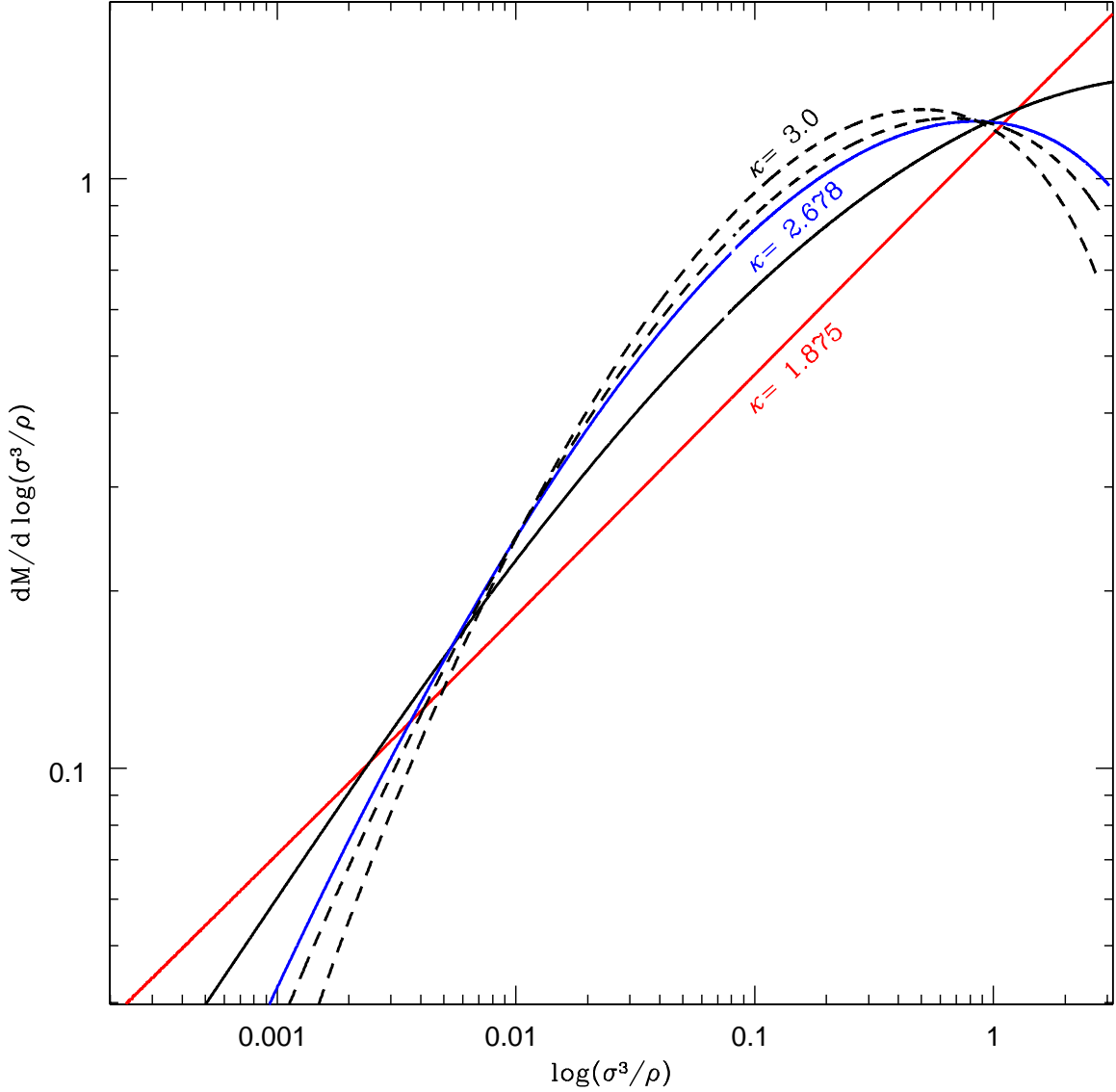}
\caption[fig4.eps]{
The phase-space distribution function of the five solutions shown in Figure 2,
normalized so that all have the same total mass, energy, and mean density. The
straight line corresponds to the power-law solution, $\kappa=1.875$. As $\kappa$
increases the distribution becomes more sharply peaked. Dashed lines correspond
to the hollow (non-physical) density profiles. The critical solution,
$\kappa=\kappa_{\rm crit}=2.678$, corresponds to the most sharply peaked
phase-space density distribution or, equivalently, to the case where phase-space
density is most uniformly distributed across the system.\label{fig4}}
\end{figure}


\begin{thebibliography}{}

%Bertschinger 1985
\bibitem[Bertschinger(1985)]{1985ApJS...58...39B} Bertschinger, E.\ 1985, 
\apjs, 58, 39 

%Binney and Tremaine 1987, p.\ 198
\bibitem[Binney \& Tremaine (1987)]{bt87} Binney, J. \& Tremaine, S. 1987, 
Galactic Dynamics (Princeton: Princeton University Press)

%Blais-Ouellette \etal 2001 
\bibitem[Blais-Ouellette \etal (2000)]{bo01}Blais-Ouellette, S., Amram, P.
\& Carignan, C. 2001, \aj, in press (astro-ph/0006449)

%Carollo \etal 1995 
\bibitem[Carollo et al.(1995)]{1995ApJ...441L..25C} Carollo, C.\ M., de 
Zeeuw, P.\ T., van der Marel, R.\ P., Danziger, I.\ J. \& Qian, E.\ E.\ 
1995, \apjl, 441, L25 

%Cole \& Lacey 1996 
\bibitem[Cole \& Lacey(1996)]{1996MNRAS.281..716C} Cole, S. \& Lacey, C.\ 
1996, \mnras, 281, 716 

%Cote \etal 2000 
\bibitem[C{\^o}t{\'e}, Carignan, \& Freeman(2000)]{2000AJ....120.3027C} 
C{\^o}t{\'e}, S., Carignan, C. \& Freeman, K.\ C.\ 2000, \aj, 120, 3027 

%Cretton \etal 2000 
\bibitem{crz} Cretton N., Rix H.-W. \& de Zeeuw P.T., 2000, ApJ, 536, 319

%Crone, Evrard \& Richstone 1994 
\bibitem[Crone, Evrard, \& Richstone(1994)]{1994ApJ...434..402C} Crone, M.\ 
M., Evrard, A.\ E. \& Richstone, D.\ O.\ 1994, \apj, 434, 402 

%de Blok \etal 2001
\bibitem[de Blok \etal (2001)]{db01}de Blok, W. J. G., McGaugh, S. S., Bosma, 
A. \& Rubin, V. C. 2001, \apjl, in press (astro-ph/0103102)

%Dubinski \& Carlberg 1991 
\bibitem[Dubinski \& Carlberg(1991)]{1991ApJ...378..496D} Dubinski, J. \& 
Carlberg, R.\ G.\ 1991, \apj, 378, 496 

%Efstathiou \etal 1985
\bibitem[Efstathiou, Davis, White, \& Frenk(1985)]{1985ApJS...57..241E} 
Efstathiou, G., Davis, M., White, S.\ D.\ M. \& Frenk, C.\ S.\ 1985, 
\apjs, 57, 241 

%Evans \& Collett 1997 
\bibitem[Evans \& Collett(1997)]{1997ApJ...480L.103E} Evans, N.\ W. \& 
Collett, J.\ L.\ 1997, \apjl, 480, L103 

%Fillmore \& Goldreich 1984 
\bibitem[Fillmore \& Goldreich(1984)]{1984ApJ...281....1F} Fillmore, J.\ 
A. \& Goldreich, P.\ 1984, \apj, 281, 1 

%Flores \& Primack 1994 
\bibitem[Flores \& Primack(1994)]{1994ApJ...427L...1F} Flores, R.\ A. \& 
Primack, J.\ R.\ 1994, \apjl, 427, L1 

%Flores \etal 1993 
\bibitem[Flores, Primack, Blumenthal, \& Faber(1993)]{1993ApJ...412..443F} 
Flores, R., Primack, J.\ R., Blumenthal, G.\ R. \& Faber, S.\ M.\ 1993, 
\apj, 412, 443 

%Frenk \etal 1988 
\bibitem[Frenk, White, Davis, \& Efstathiou(1988)]{1988ApJ...327..507F} 
Frenk, C.\ S., White, S.\ D.\ M., Davis, M., \& Efstathiou, G.\ 1988, \apj, 
327, 507 

%Fukushige \& Makino 2000 
\bibitem[Fukushige \& Makino(2000)]{fm00} Fukushige, T. \& 
Makino, J. 2000, \apj, submitted (astro-ph/0008104)

%Fukushige \& Makino 1997 
\bibitem[Fukushige \& Makino(1997)]{1997ApJ...477L...9F} Fukushige, T.\ \& 
Makino, J.\ 1997, \apjl, 477, L9 

%Gerhard \etal 1998 
\bibitem[Gerhard, Jeske, Saglia, \& Bender(1998)]{1998MNRAS.295..197G} 
Gerhard, O., Jeske, G., Saglia, R.\ P. \& Bender, R.\ 1998, \mnras, 295, 
197 

%Ghigna \etal 2000
\bibitem[Ghigna et al.(2000)]{2000ApJ...544..616G} Ghigna, S., Moore, B., 
Governato, F., Lake, G., Quinn, T. \& Stadel, J.\ 2000, \apj, 544, 616 

\bibitem[Gunn \& Gott(1972)]{1972ApJ...176....1G} Gunn, J.\ E. \& Gott, 
J.\ R.\ I.\ 1972, \apj, 176, 1 

%Hoffmann \& Shaham 1985 
\bibitem[Hoffman \& Shaham(1985)]{1985ApJ...297...16H} Hoffman, Y. \& 
Shaham, J.\ 1985, \apj, 297, 16 

%Huss, Jain \& Steinmetz 1999 
\bibitem{hjs} Huss A., Jain B. \& Steinmetz M., 1999, MNRAS, 308, 1011

%Jing \& Suto 2000
\bibitem[Jing \& Suto(2000)]{2000ApJ...529L..69J} Jing, Y.\ P. \& Suto, 
Y.\ 2000, \apjl, 529, L69 

%Klypin \etal 2000. 
\bibitem{kkbp2} Klypin A.A., Kravtsov A.V., Bullock J.S. \& Primack J.R.,
2000, ApJ, submitted (astro-ph/0006343)

%Kravtsov \etal 1998 
\bibitem{kkbp} Kravtsov A.V., Klypin A.A., Bullock J.S. \& Primack J.R.,
1998, ApJ, 502, 48

%Kronawitter \etal 2000.
\bibitem{ksgb} Kronawitter A., Saglia R.P., Gerhard O. \& Bender R.,
2000, A\&AS, 144, 53

%Lokas \& Hoffman 2000
\bibitem[{\L}okas \& Hoffman(2000)]{2000ApJ...542L.139L} {\L}okas, E.\ L. 
\& Hoffman, Y.\ 2000, \apjl, 542, L139 

%McGaugh \& de Blok 1998 
\bibitem[McGaugh \& de Blok(1998)]{1998ApJ...499...41M} McGaugh, S.\ S. \& 
de Blok, W.\ J.\ G.\ 1998, \apj, 499, 41 

%Moore 1994 
\bibitem[Moore(1994)]{1994Natur.370..629M} Moore, B.\ 1994, \nat, 370, 629 

%Moore \etal 1998
\bibitem{metal98} Moore B., Governato F., Quinn T., Stadel J. \& Lake G.,
1998, ApJ, 499, L5

%Moore \etal 1999. 
\bibitem{metal99a} Moore B., Ghigna S., Governato F., Lake G., Quinn
T., Stadel J. \& Tozzi P., 1999a, ApJ, 524, L19

%Moore \etal 1999b 
\bibitem{metal99b} Moore B., Quinn T., Governato F., Stadel J. \& Lake G.,
1999b, MNRAS, 310, 1147

%Navarro 1998 
\bibitem[]{nav98} Navarro, J.F.\ 1998, (astro-ph/9807084)

%Navarro, Frenk \& White 1996
\bibitem{nfw96} Navarro J.F., Frenk C.S. \& White S.D.M., 1996, ApJ, 462, 563

%Navarro, Frenk \& White 1997
\bibitem{nfw97} Navarro J.F., Frenk C.S. \& White S.D.M., 1997, ApJ,
490, 493

%Nusser \& Sheth 1999 
\bibitem[Nusser \& Sheth(1999)]{1999MNRAS.303..685N} Nusser, A. \& Sheth, 
R.\ K.\ 1999, \mnras, 303, 685 

%Rix \etal 1997 
\bibitem[Rix et al.(1997)]{1997ApJ...488..702R} Rix, H., de Zeeuw, P.\ T., 
Cretton, N., van der Marel, R.\ P. \& Carollo, C.\ M.\ 1997, \apj, 488, 702 

%Salmon \& Zurek 1986 
\bibitem[Quinn, Salmon, \& Zurek(1986)]{1986Natur.322..329Q} Quinn, P.\ J., 
Salmon, J.\ K. \& Zurek, W.\ H.\ 1986, \nat, 322, 329

%Swaters, Madore \& Trewhella 2000 
\bibitem{smt} Swaters R.A., Madore B.F. \& Trewhella M., 2000, ApJ, 531, L107

%Syer \& White 1998 
\bibitem[Syer \& White(1998)]{1998MNRAS.293..337S} Syer, D. \& White, S.\ 
D.\ M.\ 1998, \mnras, 293, 337 

%Tyson, Kochanski \& dell'Antonio 1998
\bibitem[Tyson, Kochanski, \& dell'Antonio(1998)]{1998ApJ...498L.107T} 
Tyson, J.\ A., Kochanski, G.\ P. \& dell'Antonio, I.\ P.\ 1998, \apjl, 
498, L107 

%van den Bosch \& Swaters 2000
\bibitem{vdbs} van den Bosch F.C. \& Swaters R.A., 2000, AJ, submitted,
(astro-ph/0006048)

%van den Bosch \etal 2000 
\bibitem{vdbea} van den Bosch F.C., Robertson B.E., Dalcanton J.J. \& de
Blok W.J.G., 2000, AJ, 119, 1579

%Williams, Navarro \& Bartelmann 1999
\bibitem[Williams, Navarro \& Bartelmann(1999)]{1999ApJ...527..535W} 
Williams, L.\ L.\ R., Navarro, J.\ F., \& Bartelmann, M.\ 1999, \apj, 527, 
535 

%Zaritsky \& White 199?
\bibitem[White \& Zaritsky(1992)]{1992ApJ...394....1W} White, S.\ D.\ M.\ 
\& Zaritsky, D.\ 1992, \apj, 394, 1

\end{thebibliography}
\end{document}